\documentstyle[12pt]{article}

\newcommand{\beq}{\begin{equation}}
\newcommand{\eeq}{\end{equation}}
\newcommand{\beqa}{\begin{eqnarray}}
\newcommand{\eeqa}{\end{eqnarray}}

\newcommand{\eqn}[1]{(\ref{#1})}
\newcommand{\eq}[1]{eq.\ (\ref{#1})}

\newcommand{\sfrac}[2]{{\textstyle \frac{#1}{#2}}}
\newcommand{\n}{\nonumber \\}
\newcommand{\ch}{{\rm ch}}
\font\csc=cmcsc10 scaled\magstep1
\newcommand{\Winf}{W_{1+\infty}}
\newcommand{\Win}{W_{\infty}}
\newcommand{\Winfp}{W_{1+\infty}[p(D)]}
\newcommand{\tW}{\widetilde{W}}
\newcommand{\tD}{\widetilde{\Delta}}
\newcommand{\glinf}{\hat{\mbox{gl}}(\infty)}
\newcommand{\glinfp}{\glinf(\left\{ s \right\})}
\newcommand{\bt}{\tilde b}

\font\tennn=msbm10
\font\twelvenn=msbm10 scaled\magstep1

\newcommand{\Sbm}[1]{\leavevmode\raise-.25ex\hbox{\twelvenn #1}}
\newcommand{\sbm}[1]{\leavevmode\raise-.15ex\hbox{\tennn #1}}
\newcommand{\bC}{{\Sbm C}}

\newcommand{\bZ}{{\Sbm Z}}
\newcommand{\bz}{{\sbm Z}}

\begin{document}

{\baselineskip=14pt
 \rightline{
 \vbox{\hbox{RIMS-985}
       \hbox{YITP/K-1076}
       \hbox{YITP/U-94-22}
       \hbox{SULDP-1994-4}
       \hbox{June 1994}
}}}

\vskip 12mm
\begin{center}
{\large\bf
Subalgebras of $\Winf$ and Their Quasifinite Representations\\
}

\vspace{15mm}

\renewcommand{\thefootnote}{\fnsymbol{footnote}}
{\csc Hidetoshi AWATA}\footnote{JSPS fellow}\setcounter{footnote}
{0}\renewcommand{\thefootnote}{\arabic{footnote}}\footnote{
      e-mail address : awata@kurims.kyoto-u.ac.jp},
{\csc Masafumi FUKUMA}\footnote{
      e-mail address : fukuma@yukawa.kyoto-u.ac.jp} ,
{\csc Yutaka   MATSUO}\footnote{
      e-mail address : yutaka@yukawa.kyoto-u.ac.jp}\\
\vskip.1in
and \
{\csc Satoru    ODAKE}\footnote{
      e-mail address : odake@jpnyitp.yukawa.kyoto-u.ac.jp}

{\baselineskip=16pt
\it\vskip.25in
  $^1$Research Institute for Mathematical Sciences \\
  Kyoto University, Kyoto 606, Japan \\
\vskip.1in
  $^2$Yukawa Institute for Theoretical Physics \\
  Kyoto University, Kyoto 606, Japan \\
\vskip.1in
  $^3$Uji Research Center, Yukawa Institute for Theoretical Physics \\
  Kyoto University, Uji 611, Japan \\
\vskip.1in
  $^4$Department of Physics, Faculty of Liberal Arts \\
  Shinshu University, Matsumoto 390, Japan
}

\end{center}

\vspace{12mm}

\begin{abstract}

We propose a series of new subalgebras of the $\Winf$ algebra
parametrized by polynomials $p(w)$, and study their quasifinite
representations.
We also investigate the relation between such subalgebras and
the $\glinf$ algebra.
As an example, we investigate the $\Win$ algebra which corresponds
to the case $p(w)=w$, presenting its free field realizations and Kac
determinants at lower levels.

\end{abstract}

\vskip.3in
hep-th/9406111
\setcounter{footnote}{0}
\newpage
\section{Introduction}

It is known that there are several types of $W$ infinity algebras,
including the $w_{\infty}$ algebra as the algebra of area preserving
diffeomorphism of two-dimensional cylinder \cite{rB},
the $\Win$ algebra as its deformation $\cite{rPRS1}$, and
the $\Winf$ algebra by adding a spin-1 current to $\Win$ \cite{rPRS2}.
The $\Winf$ algebra may be regarded as the most fundamental
because all other $W$ infinity algebras are obtained as its
subalgebras \cite{rPRS3}.

The representation theories of these algebras had not been
developed after the early works \cite{rBK,rO}, because of the difficulty
that there might exist infinitely many states at each energy level
reflecting the infinite number of currents.
The requirement that there exist only a finite number of states at
each energy level is quite natural from the view point of the free-field
realization, since such quasifiniteness condition is actually ensured
when states are generated by (a finite number of) free-field
oscillators.
Recently Kac and Radul gave an elegant framework to set up the
quasifinite condition for the $\Winf$ algebra, and studied such
representations in detail \cite{rKR}.
On the basis of their analysis, further studies were made
for the $\Winf$ algebra \cite{rM,rAFMO1,rAFMO3,rFKRW} and
the super $\Winf$ algebra \cite{rAFMO2}.

In this letter, we propose a systematic method to construct
a family of subalgebras of the $\Winf$ algebra, and then study their
quasifinite representations.
In particular, we investigate the $\Win$ algebra as a special case,
and present the free field realizations and
the Kac determinant formula at lower levels.

\section{Subalgebras of $\Winf$}

The $\Winf$ algebra is defined as a one-dimensional central
extension of the Lie algebra of differential operators on the
circle whose classical generators are $z^n D^k$ ($D\equiv z
\frac{d}{dz}$, $n \in \bZ$, $k\in \bZ_{\geq 0}$).
We denote the corresponding generators in the $\Winf$ algebra by
$W(z^n D^k)$ and the central charge by $C$.
The commutation relations are defined by \cite{rKR}
\beqa
  &&\left[ W (z^n f(D)), W (z^m g(D))\right] \n
  &\!\!=\!\!&
  W (z^{n+m}f(D+m)g(D))
  -W (z^{n+m}f(D)g(D+n)) \n
  && +C \delta_{n+m,0}
  \left( \theta(n\geq 1) \sum_{j=1}^n  f(-j) g(n-j)
        -\theta(m\geq 1) \sum_{j=1}^m f(m-j) g(-j)
  \right),
\eeqa
where $\theta(P)=1$ (or 0) when the proposition $P$ is true (or false).

In the above expression, $f(D)$ and $g(D)$ are arbitrary polynomials.
However, the commutation relations still close even for a class of
the polynomials which can be divided by a polynomial $p(D)$.
In fact, if we set
\beq
  \tW(z^n D^k)=W(z^n D^k p(D)),
\eeq
then their commutation relations will be written as
\beqa
  &&\left[ \tW (z^n f(D)), \tW (z^m g(D))\right] \n
  &\!\!=\!\!&
  \tW (z^{n+m} p(D+m)f(D+m)g(D))
  -\tW (z^{n+m} p(D+n)f(D)g(D+n)) \n
  && +C \delta_{n+m,0}
  \biggl( \theta(n\geq 1) \sum_{j=1}^n p(-j)p(n-j) f(-j) g(n-j) \n
 && \hspace{20mm}
        -\theta(m\geq 1) \sum_{j=1}^m p(-j)p(m-j) f(m-j) g(-j)
  \biggr).
\eeqa
We call the obtained subalgebra $\Winfp$.
The motivation for our construction of such subalgebras is as follows.
In the $\Win$ algebra, there is no spin-1 current.
Since the current is expressed as $W(z^n)$ in $\Winf$,
the $\Win$ algebra will be obtained from $\Winf$ simply by omitting
this current.
However, it is equivalent to taking $p(D)=D$ in the above expression.

By introducing $z^n e^{xD}$ as a generating series for $z^n D^k$,
the above commutation relation can be rewritten in a simpler form:
\beqa
  \left[ \tW(z^n e^{xD}), \tW(z^m e^{yD})\right]
  &\!\!=\!\!&
  \left( p(\sfrac{d}{dx}) e^{mx}
         -p(\sfrac{d}{dy}) e^{ny} \right)
   \tW(z^{n+m}e^{(x+y)D}) \n
  &&
  -C p(\sfrac{d}{dx}) p(\sfrac{d}{dy})
  \frac{e^{mx}-e^{ny}}{e^{x+y}-1}\delta_{n+m,0}.
\eeqa

The basis of $\Winf$ given in \cite{rPRS2}, $V^i_n=W^{i+2}_n$, is
expressed as $W^{k+1}_n=W(z^n f^k_n(D))$ ($k\geq 0$),
\beq
  f^k_n(D)={2k \choose k}^{-1}
  \sum_{j=0}^k(-1)^j{k \choose j}^2
  \lbrack -D-n-1\rbrack_{k-j} \lbrack D\rbrack_j,
  \label{fkn}
\eeq
where $\lbrack x\rbrack_n=\prod_{j=0}^{n-1}(x-j)$ and
${x \choose n}=\lbrack x\rbrack_n/n!$.
On the other hand, the basis of $\Win$ given in \cite{rPRS1},
$\widetilde{V}^i_n=\tW^{i+2}_n$, is now expressed as
$\tW^{k+2}_n=\tW (z^n \tilde{f}^k_n(D))$ ($k\geq 0$),
\beq
  \tilde{f}^k_n(D)=-{2(k+1) \choose k+1}^{-1}
  \sum_{j=0}^k(-1)^j {k \choose j} {k+2 \choose j+1}
  \lbrack -D-n-1 \rbrack_{k-j} \lbrack D-1 \rbrack_j.
  \label{ftkn}
\eeq

We remark that the Virasoro generators exist only if $\deg p(w) \leq 1$.
In the case of $\Win$, the Virasoro generator $L_n$ is given
by $L_n=-\tW (z^n)$ whose central charge, $\tilde{c}_{Vir}$, is
related to $C$ as $\tilde{c}_{Vir}=-2C$ \cite{rPRS3}.
For $\deg p(w) \geq 2$, we extend the algebra introducing the $L_0$
operator such as to count the energy level,
$\left[ L_0, \tW (z^n f(D)) \right]$ $=$ $-n\tW (z^n f(D))$.

\section{Quasifinite Representations}

We consider the irreducible quasifinite
highest weight representations of $\Winfp$, following \cite{rKR}.

The highest weight state $|\lambda\rangle$ is characterized by
\beqa
  \tW (z^nD^k) |\lambda\rangle
  &\!\!=\!\!& 0 \qquad\quad (n \geq 1, k \geq 0), \n
  \tW (D^k) |\lambda\rangle
  &\!\!=\!\!& \tD_k |\lambda\rangle \quad (k \geq 0).
\eeqa
It is convenient to introduce the generating function for the
highest weights $\tD_k$,
\beq
  \tD (x)=-\sum_{k=0}^{\infty} \tD_k \frac{x^k}{k!},
\eeq
which is the eigenvalue of $-\tW (e^{xD})$;
$\tW (e^{xD}) |\lambda\rangle=-\tD (x) |\lambda\rangle$.
By definition, $\tD(x)$ must be regular at $x=0$.
To study the representation of $\Winfp$, we first consider the
representation of $\Winf$ whose restriction agrees with the
representation of $\Winfp$. Such a representation of $\Winf$ always
exists.
The weight $\Delta(x)$, which is defined by
$W(e^{xD})|\lambda\rangle = - \Delta(x) |\lambda\rangle$, satisfies
\beq
 p \left( \sfrac{d}{dx} \right) \Delta(x) = \tD(x).
\eeq
The representation of the $\Winf$ associated with that of
$\Winfp$ is not uniquely determined,
and we will fix one representative $\Delta(x)$.

The quasifinite representation of $\Winf$ with $\Delta(x)$ is
characterized by characteristic polynomials $b_n(w)$
($n=1,2,3,\cdots$) \cite{rKR} (see also \cite{rAFMO3}):
\beq
  b_n(w)={\rm lcm}(b(w),b(w-1),\cdots,b(w-n+1)),
\eeq
where $b(w)=b_1(w)$ is the minimal-degree monic polynomial satisfying
the differential equation,
\beq
  b \left( \sfrac{d}{dx} \right)
  \Bigl((e^x-1)\Delta(x)+C\Bigr) = 0.
\eeq

Quasifinite condition is the requirement that there exist a finite
number of states at each energy level.
This means that $\tW (z^{-n}f(D))|\lambda\rangle$ is a null state
for sufficiently high-degree polynomial $f(w)$.
In other words, the set $\tilde{I}_{-n}=\left\{f(w)\in\bC[w]|
\tW(z^{-n}f(D))|\lambda\rangle
=
\mbox{null state} \right\}$ contains
an ideal of $\bC[w]$ generated by some polynomial.
We can show that if there exist a finite number of states at level 1,
then this is also the case at any level.
Suppose that $\tW(z^{-1}f(D))|\lambda\rangle$ is a null state, then
\beqa
  0&\!\!=\!\!&
  \tW(ze^{x(D+1)})\tW(z^{-1}f(D))|\lambda\rangle \n
  &\!\!=\!\!&
  f\left(\sfrac{d}{dx}\right)
  \biggl(\left(p(\sfrac{d}{dx})e^x-p(\sfrac{d}{dx}-1)\right)\tD(x)
  +Cp(0)p(-1)\biggr)|\lambda\rangle \n
  &\!\!=\!\!&
  f\left(\sfrac{d}{dx}\right)
  p\left(\sfrac{d}{dx}\right)
  p\left(\sfrac{d}{dx}-1\right)
  \Bigl((e^x-1)\Delta(x)+C\Bigr)|\lambda\rangle.
\eeqa
Therefore, $\Delta(x)$ is the weight for the quasifinite representation.

For $\deg p(w)\leq 2$, we can prove that $\tilde{I}_{-n}$ is an
ideal of $\bC[w]$, as is the case for $\Winf$.
However, this is no longer true when $\deg p(w)\geq 3$.
In the following, we make an assumption that $\tilde{I}_{-n}$ is
an ideal of $\bC[w]$.
Since $\bC [w]$ is a principal ideal domain, $\tilde{I}_{-n}$ is
generated by a monic polynomial $\tilde{b}_n(w)$.
We call the $\tilde{b}_n(w)$'s $(n=1,2,3,...)$ the characteristic
polynomials for the highest weight representation.

The characteristic polynomials $\tilde{b}_n(w)$ are related to each
other as follows:\footnote{
We can also show that $\tilde{b}_{n+m}(w)$ divides
$p(w-m)\tilde{b}_n(w-m)\tilde{b}_m(w)$.}\\
{}~~~~(\romannumeral1)
  $\tilde{b}_n(w)$ divides both of $p(w+m)\tilde{b}_{n+m}(w+m)$
  and $p(w-n-m)\tilde{b}_{n+m}(w)~{~}^\forall m\in\bZ_{\geq 0},$ \\
{}~~~~(\romannumeral2) $p(w)p(w-n)\tilde{b}_n(w)$ is divided
  by $f_n(w)$.\\
The $\tilde{b}_n(w)$'s are determined as minimal-degree polynomials
satisfying (\romannumeral1) and (\romannumeral2).
Here $f_n(w)$ is the minimal-degree, monic polynomial satisfying
the following differential equation:
\beq
  f_n \left( \sfrac{d}{dx} \right)
  \sum_{j=0}^{n-1} e^{jx}
  \Bigl( (e^x-1)\Delta(x)+C \Bigr) = 0.
\eeq
Note that $f_1(w)=b(w)$, and $b_n(w)$ is divided by $f_n(w)$.
The property (\romannumeral1) is derived from the null state condition,
$\tW(z^me^{xD})\tW(z^{-n-m}\tilde{b}_{n+m}(w))|\lambda\rangle=$
null state.
The property (\romannumeral2) is derived from the following null state
condition,
\beqa
  0&\!\!=\!\!&
  \tW(z^ne^{x(D+n)})\tW(z^{-n}\tilde{b}(D))|\lambda\rangle \n
  &\!\!=\!\!&
  \tilde{b}\left(\sfrac{d}{dx}\right)
  p\left(\sfrac{d}{dx}\right)
  p\left(\sfrac{d}{dx}-n\right)
  \sum_{j=0}^{n-1} e^{jx}
  \Bigl( (e^x-1)\Delta(x)+C \Bigr)|\lambda\rangle.
\eeqa

In almost all cases, the relation $f_n(w)=b_n(w)$ holds.
Then, we can prove that the characteristic polynomials
$\tilde{b}_n(w)$ are given by
\beq
  \tilde{b}_n(w)=\frac{b_n(w)}{\mbox{gcd}(b_n(w),p(w)p(w-n))}.
  \label{tb}
\eeq
Although \eq{tb} satisfies both (\romannumeral1) and (\romannumeral2)
in general, it might not be minimal-degree polynomials.
Therefore, for such special cases, we must solve (\romannumeral1) and
(\romannumeral2) directly.
In contrast to $\Winf$, the $\tilde{b}_n(w)$'s may not be determined by
$\tilde{b}_1(w)$ (and $p(w)$) alone.

\section{Correspondence with $\glinf$}
As was demonstrated in \cite{rKR,rAFMO3},
the quasifinite representation of $\Win$ can
be alternatively expressed in terms of $\glinf$ algebra.
Since $\glinf$ has a simpler structure,
it is useful to carry out some explicit calculation
by using this correspondence.
As we show in this section, it is also useful
to understand the structure of subalgebras
in terms of $\glinf$.

Let us start from the special case in which
the differences of any   roots of $\bt(w)=0$
and $p(w)=0$ are integers.
We assume that $\bt(w)$ has the following form:
\beq
\bt(w) = \prod_{i=1}^N (w-\lambda-k_i)^{\mu_i}
\prod_{j=1}^M (w-\lambda-s_j)^{\nu_j},\quad
p(w) = \prod_{j=1}^M (w-\lambda-s_j)^{\rho_j},
\nonumber
\eeq
where $k_i,\ s_j\in \bZ$ and $\mu_i, \nu_j, \rho_j$ are
positive integers with $m\equiv \mbox{Max}(\mu_i, \nu_j+\rho_j)$.
We remark that general cases can also be handled by making their
tensor products.

The $\glinf$ algebra we consider is defined by the commutation
relation,
\beqa
\left[ E^{(\mu)}(i,j), E^{(\nu)}(k,\ell)\right]
&=&\theta(m-\mu-\nu\geq0)\times\n
&&\left(\delta_{j+k,0}E^{(\mu+\nu)}(i,\ell)-
\delta_{i+\ell,0}E^{(\mu+\nu)}(k,j)
\right.\n
&&\left. +c^{(\mu+\nu)}\delta_{j+k,0}\delta_{\ell+i,0}
(\theta(i\geq 0)-\theta(k\geq 0))\right),
\label{glcom}
\eeqa
where $\mu$ and $\nu$ run from 0 to $m$.
The quasifinite representation of $\glinf$
is defined by the highest weight state,
\beq
E^{(\mu)}(i,j) |\lambda \rangle =
0 \quad (i+j>0)\quad \mbox{and} \quad
E^{(\mu)}(i,-i) |\lambda \rangle =
q^{(\mu)}_i |\lambda \rangle.
\nonumber
\eeq
If we introduce
$
h^{(\mu)}_i = q^{(\mu)}_i-q^{(\mu)}_{i-1}
+c^{(\mu)}\delta_{i,0},
$
the quasifiniteness of the module is achieved only when
finite number of $h^{(\mu)}_i$ are non-vanishing \cite{rKR}.

The relation between the generators of $\glinf$ and $\Winf$
is given by,
\beq
W(z^r f(D))= \sum_{k\in \bZ}\sum_{\mu=0}^m
\frac{f^{(\mu)}(\lambda +k )}{\mu !}
E^{(\mu)}(r+k,-k),
\label{wgl}
\eeq
for $r\neq 0$ \cite{rKR}.
For zero modes, we need to introduce c-number corrections
\cite{rKR,rAFMO3}.

Since we are considering a subalgebra of $\Winf$,
we need to find a proper subalgebra of $\glinf$
associated with it.
Let us replace $f(D)$ in \eqn{wgl} by $\tilde{f}(D) p(D)$.
We recognize that the coefficients of
\beq
E^{(\mu)}(r, -s_j)\qquad (r\in\bZ,~~\mu=0,\cdots, \rho_j\!-\!1),
\label{dead}
\eeq
vanish on the right hand side of \eqn{wgl}.
This means that these generators will not
appear in the image of the mapping \eqn{wgl}.
Thus, if we remove \eqn{dead} from the algebra,
we also have to remove the generators of the following form:\footnote{
To prove this, we first remark that generators
of the form, $E^{(\nu)}(*,-s_j)$ with $\nu=\rho_j,\cdots, m$,
still remain in the algebra.  The generators which has
non-vanishing inner product with them are given in the form,
$E^{(\nu)}(s_j,*)$ with $\nu=0,\cdots,m\!-\!\rho_j$.
The other generators \eqn{dead2}
have zero inner product with any states.
}
\beq
E^{(\mu)}(s_j,r)\qquad (r\in\bZ,~~\mu=m\!-\!\rho_j\!+\!1,\cdots, m).
\label{dead2}
\eeq
The remaining generators will form a subalgebra of $\glinf$, which
will be called $\glinfp$ in the following.

The next step is to find a relation between the quasifinite
representations of $\Winfp$ and $\glinfp$.
This is again carried out by studying the relation
\eqn{wgl} carefully.
By putting $f(D)=\tilde{b}(D)p(D)$, $r=-1$, it
should become a null field when acting on the highest weight state.
The only generators which may give nonvanishing states
are,
$
E^{(\mu)}(-1+k_i,-k_i)$ $(\mu=0,\cdots,\mu_i-1)
$ and
$
E^{(\mu)}(-1+s_j,-s_j)$ $(\mu=\rho_j,\cdots,\rho_j+\nu_j-1).
$
Thus, applying the argument in \cite{rKR}, we can
show that nonvanishing $h^{(\mu)}$ are given by
$
h^{(\mu)}_{k_i}~~(\mu=0,\cdots,\mu_i-1)
$ and
$
h^{(\mu)}_{s_j}~~(\mu=\rho_j,\cdots,\rho_j+\nu_j-1).
$
This information is enough for constructing a $\glinfp$ module
from the highest weight state, which
can be shown to be equivalent to the $\Winfp$ module
we gave in section 3.

\section{Free Field Realization of the $\Win$ Algebra}

In this section, we investigate the $\Win$ algebra $(p(D)=D)$ as an
example.

We first consider the free-field realization of $\Win$, introducing
complex bosons,
$\bar{\varphi}(z)=\bar{q} + \bar{\alpha}_0 \log z$
$-\sum_{n \neq 0}\frac{1}{n}\bar{\alpha}_n z^{-n}$,
$\varphi(z)=q + \alpha_0 \log z$
$-\sum_{n \neq 0}\frac{1}{n}\alpha_n z^{-n}$
with commutation relations:
$\left[ \bar{\alpha}_n, \alpha_m \right]=n\delta_{n+m,0}$,
$\left[ \bar{\alpha}_0, q \right]=1$,
$\left[ \alpha_0, \bar{q} \right]=1$,
{\it i.e.}, with the propagator $\bar{\varphi}(z) \varphi(w) \sim
\log(z-w)$.
The $\Win$ algebra with $C=-1$ ($\widetilde{c}_{Vir}=2$) is then
realized in the following way \cite{rBK}:
\beq
  \tW (z^n f(D)) = -\oint \frac{dz}{2\pi i}
  :\partial \bar{\varphi}(z) z^{n+1} f(D+1) \partial \varphi(z):~
  = -\sum_{m \in \bz} f(-m) :\bar{\alpha}_{n-m} \alpha_{m}:.
\eeq
We remark that
\beq
  \left[ \tW(z^n f(D)),\bar{\alpha}_m \right]=mf(m)\bar{\alpha}_{n+m},
  \quad
  \left[ \tW(z^n f(D)),\alpha_m \right]=mf(-n-m)\alpha_{n+m}.
  \label{tWalpha}
\eeq

There are two kinds of highest weight states for such
realization \cite{rBK,rO}:
\beqa
  |p,\bar{p}\rangle &\!\!=\!\!&
  :\exp \left( p\bar{\varphi}(0) + \bar{p}\varphi(0) \right):
  |0\rangle, \\
  |N\rangle &\!\!=\!\!&
  \left\{
  \begin{array}{ll}
    \left( \alpha_{-1} \right)^N |0\rangle & (N \geq 1) \\
    |0\rangle & (N = 0) \\
    \left( \bar{\alpha}_{-1} \right)^{-N} |0\rangle & (N \leq -1),
  \end{array}
  \right.
\eeqa
where $p,\bar{p} \in \bC$, $N \in \bZ$ and $|0\rangle$ is defined by
$\bar{\alpha}_n |0\rangle=\alpha_n |0\rangle=0$ ($n \geq 0$).
Using \eq{tWalpha},
we can show that the weight $\tD(x)$ and the characteristic
polynomials $\tilde{b}_n(w)$ are given by the following:
for $|p,\bar{p}\rangle$
\beqa
  \tD(x) &\!\!=\!\!&
  p\bar{p}, \\
  \tilde{b}_n(w) &\!\!=\!\!&
  \left\{
  \begin{array}{lll}
    \lbrack w-1 \rbrack_{n-1}& \mbox{if }& p\bar{p}=0 \\
    \lbrack w   \rbrack_{n+1}& \mbox{if }& p\bar{p}\neq 0,
  \end{array}
  \right.
\eeqa
and for $|N\rangle$
\beqa
  \tD(x) &\!\!=\!\!&
  |N| e^{sgn(N) x}, \\
  \tilde{b}_n(w) &\!\!=\!\!&
  \left\{
  \begin{array}{llll}
    \lbrack w-1 \rbrack_{n-1} \cdot (w-n-1)& \mbox{if }& N \geq 1 \\
    \lbrack w-1 \rbrack_{n-1}              & \mbox{if }& N = 0 \\
    \lbrack w-1 \rbrack_{n-1} \cdot (w+1)  & \mbox{if }& N \leq -1,
  \end{array}
  \right.
\eeqa

We can illustrate our general theory given in the previous section
from this example.
For $|p,\bar{p}\rangle$, we can take $\Delta(x)=p\bar{p}x$. Then
$f_n(w)$ is equal to $b_n(w)=[w]_n$ for $p\bar{p}=0$.
Thus $\tilde{b}_n(w)$ are given by \eq{tb}.
But for $p\bar{p}\neq 0$, $f_n(w)=[w]_{n+1}w(w-n)$ does not agree
with $b_n(w)=([w]_{n+1})^2$. So the polynomials in \eq{tb} may not be
of minimal degree, and actually they are not.

Using \eq{tWalpha} and a formula in \cite{rO,rAFOQ},
we can write down the full character of the representation
$|p,\bar{p}\rangle$ with $p\bar{p} \neq 0$ as
\beq
  \ch_{p,\bar{p}}=\mbox{tr}
  e^{\sum_{k=0}^{\infty} g_k \tW(D^k)}
  =
  e^{-p\bar{p}g_0}
  \prod_{n=1}^{\infty}
  \left( 1-e^{\sum_{k=0}^{\infty} g_k n^{k+1}} \right)^{-1}
  \left( 1-e^{-\sum_{k=0}^{\infty} g_k (-n)^{k+1}} \right)^{-1},
\eeq
and the generating function of the full characters of the
representations $|N\rangle$,
$\ch_N=\mbox{tr} e^{\sum_{k=0}^{\infty} g_k \tW(D^k)}
=e^{-\sum_{k=0}^{\infty}|N|(sgn(N))^kg_k}\ch_N'$, as
\beq
  \sum_{N \in \bz} t^N
  \ch_N'
  =
  \prod_{n=1}^{\infty}
  \left( 1-t^{-1} e^{\sum_{k=0}^{\infty} g_k n^{k+1}} \right)^{-1}
  \left( 1-t e^{-\sum_{k=0}^{\infty} g_k (-n)^{k+1}} \right)^{-1}.
\eeq

We here make a comment.
The $\Winf$ algebra is known to be realized by free fermions or ${\bf
bc}$ ghosts \cite{rBPRSS,rM}.
By restricting quasifinite representations of $\Winf$, we obtain
those of $\Win$ or more generally $\Winfp$.

\section{Discussion}

We have constructed a series of subalgebras of $\Winf$ parametrized
by polynomials $p(w)$, in which $\Win$ corresponds to $p(w)=w$,
and studied their quasifinite representations.

Although $\Win$ is a subalgebra of $\Winf$, its representation
theory is nontrivial. There exist more null states than the $\Winf$
case. Although full character formulae of quasifinite representations
of $\Winf$ were recently obtained \cite{rAFMO3,rFKRW},
it is difficult to derive the (full) character formulae of
$\Win$ directly from those of $\Winf$.
We would like to report on this issue in our future communication.

\vskip 5mm
\noindent{\bf Acknowledgments:}
S.O. would like
to thank members of YITP for their hospitality.
This work is supported in part by
Grant-in-Aid for Scientific Research from Ministry of
Science and Culture.

\newcommand{\la}{\lambda}
\section* {Appendix: Kac Determinant at Lower Degrees}
\setcounter{section}{1}
\renewcommand{\thesection}{\Alph{section}}
\setcounter{equation}{0}

In this appendix, we give the Kac determinant for the $p(w)=w$ and
$\tilde b(w)= w-\la$ case.
In this case, the generating function for the highest weight can be
written as $\Delta(x)= C_1(e^{\la x}-1)/(e^x-1)$
with $C_1$ some complex number.

For the first three levels, the relevant ket states are,
\beqa
\mbox{Level 1}&\quad&
\tW(z^{-1})|\lambda\rangle,\n
\mbox{Level 2}&\quad&
\tW(z^{-2})|\lambda\rangle,\
\tW(z^{-2}D)|\lambda\rangle,\
\tW(z^{-2}D^2)|\lambda\rangle,\
\tW(z^{-1})^2|\lambda\rangle,\n
\mbox{Level 3}&\quad&
\tW(z^{-3})|\lambda\rangle,\
\tW(z^{-3}D)|\lambda\rangle,\
\tW(z^{-3}D^2)|\lambda\rangle,\
\tW(z^{-3}D^3)|\lambda\rangle,\
\tW(z^{-3}D^4)|\lambda\rangle,\n
&\quad&
\tW(z^{-1})\tW(z^{-2})|\lambda\rangle,\
\tW(z^{-1})\tW(z^{-2}D)|\lambda\rangle,\n
&\quad&
\tW(z^{-1})\tW(z^{-2}D^2)|\lambda\rangle,\
\tW(z^{-1})^3|\lambda\rangle. \nonumber
\eeqa
Corresponding bra states may be given by changing
$z^{-r}$ into $z^r$.

We compute the Kac determinant up to level 4, and the result is
\beqa
\det[1]&\propto& C_1\la (\la-1),\n
\det[2]&\propto& C_1^3 (C_1+1)  C_2
    (\la+1) \la^5 (\la-1)^5 (\la-2) ,\n
\det[3]&\propto&  C_1^7  (C_1+1)^3  (C_1-2) C_2^3
 (\la+2) (\la+1)^4\la^{14} (\la-1)^{14} (\la-2)^4 (\la-3),\n
\det[4]&\propto& (C_1+1)C_1^{18}(C_1-1)^9(C_1-2)^3(C_1-3)
C_2^9(C_2-1)\n
&\times& (\la+3) (\la+2)^4 (\la+1)^{13}\la^{42}
(\la-1)^{42} (\la-2)^{13} (\la-3)^4 (\la-4) ,\nonumber
\eeqa
where $C_2\equiv C-C_1$.


\end{document}